\documentclass[amsmath,amssymb,onecolumn,superscriptaddress,prstper,runinaddress,floatfix]{revtex4}

\usepackage{graphicx}%
\usepackage{dcolumn}%
\def\be{\begin{equation}}
\def\ee{\end{equation}}
\def\beq{\begin{eqnarray}}
\def\eeq{\end{eqnarray}}

\usepackage{bm}

\usepackage{bm}
\usepackage{amsfonts}
\usepackage{latexsym}
\usepackage{graphicx}
\usepackage{amsmath}
\usepackage{palatino}
\usepackage{mathpazo}
\usepackage{textcomp}
\usepackage{amsfonts}
\usepackage{amssymb}
\usepackage{amsmath}
\usepackage{epsfig}
\usepackage{float}
\usepackage{subfig}
\usepackage{booktabs}
\usepackage{array}
\usepackage{mathrsfs}
\usepackage{xcolor}

\allowdisplaybreaks
\begin{document}
\title{Gauss-Bonnet entropy and thermal dynamics of RN-AdS black holes}

\author{M. Z. Bhatti}
\email[Email: ]{mzaeem.math@pu.edu.pk}
\affiliation{Institute of Mathematics, University of the Punjab, Lahore-54590, Pakistan.}
\affiliation{Research Center of Astrophysics and Cosmology, Khazar University, 41 Mehseti Street, 1096, Baku, Azerbaijan.}

\author{Kazuharu Bamba}
\email[Email: ]{bamba@sss.fukushima-u.ac.jp (corresponding~author)}
\affiliation{Faculty of Symbiotic Systems Science,
Fukushima University, Fukushima 960-1296, Japan.}

\author{I. Siddique}
\email[Email: ]{iqra123.siddique123@gmail.com, iqra.siddique@ucp.edu.pk}
\affiliation{Institute of Mathematics, University of the Punjab, Lahore-54590, Pakistan.}

\author{Bander Almutairi}
\email[Email: ]{baalmutairi@ksu.edu.sa}
\affiliation{Department of Mathematics, College of Science, King Saud University, P.O. Box 2455 Riyadh 11451, Saudi Arabia.}

\author{Z. Yousaf}
\email[Email: ]{zeeshan.math@pu.edu.pk}
\affiliation{Institute of Mathematics, University of the Punjab, Lahore-54590, Pakistan.}


\begin{abstract}
We explore the thermodynamics of a novel solution for the Reissner-Nordstr\"{o}m-Anti-de Sitter (AdS) black hole, uniquely incorporating the Gauss-Bonnet term. Unlike previous studies that primarily focused on standard General Relativity or other modifications, this inclusion allows for a modified entropy formulation, facilitating the computation of key thermodynamic quantities such as Gibbs free energy, the first law of thermodynamics,  the equation of state, and Hawking temperature. We identify critical points and graphically represent the relationship between temperature and Gibbs free energy as a function of the horizon radius. Ultimately, we assess the thermal stability of the Reissner-Nordstr\"{o}m-AdS black hole within the framework of Gauss-Bonnet gravity, emphasizing the influence of the Gauss-Bonnet term unlike previous studies that primarily focused on standard General Relativity or other modifications. As a result, it is found that the Gauss-Bonnet coupling significantly alters the thermodynamic behavior and stability structure of the black hole, revealing richer phase transition phenomena. \\
\end{abstract}
\maketitle

\section{Introduction}

Black holes (BH) are regions of space with a high mass density, resulting in a strong gravitational force that prevents light from escaping. As a result, matter entering a BH is irretrievably lost. These cosmic entities often form from the remnants of massive stars that have exhausted their nuclear fuel, collapsing under their gravity. Despite being invisible to the naked eye due to their negligible light emission, BH can grow substantially over time by absorbing gas, stars, planets, and other BH. A BH has two main features: the singularity and the event horizon. The event horizon, known as the BH surface, represents the point beyond which escape is impossible due to tremendous gravitational forces. The singularity, an indefinitely small and dense point, is located at the point of center of a BH. Although, the theory proposed by Einstein supports the existence of singularities.

The term ``black hole," a captivating concept in the realm of astrophysics, was first introduced by the Ann Ewing in a magazine article in 1964, following its mention in an earlier report in the Science News Letter. However, it was the renowned physicist John Wheeler who truly popularized the term during a landmark lecture in New York in 1967, leading to its widespread incorporation into scientific discourse and popular culture. The Event Horizon Telescope project marked a historic achievement in 2019 by capturing the first-ever photograph of a BH, an image that stunned the world and opened a new chapter in our understanding of the universe. Another amazing achievement came in 2022 after this ground-breaking work: a picture of the super-massive BH that sits at the heart of the Milky Way, our galaxy. These breathtaking images enthralled viewers everywhere, stimulating their imaginations and deepening their interest in the universe. By the late 1960s, physicists had started to realize that the three basic characteristics of mass, electric charge, and angular momentum could adequately characterize the astounding features of BH, which were based on the ideas of general relativity. We may investigate their mysterious presence in the vast universe because these three characteristics are remarkably adequate to describe the complicated nature of BH.
After introducing Einstein's general relativity \cite{1}, Schwarzschild \cite{2}, Reissner-Nordstr\"{o}m \cite{3,4}, Kerr and Ezra Newman developed BH solutions that are spherically symmetric, non-rotating (with an electromagnetic field), uncharged and rotating BH (without an electromagnetic field) \cite{5}, both rotating and charged \cite{6} respectively. In addition, research on BH thermodynamics, particularly the Van der Waals (VdW) fluid method. It is a most popular topic in theoretical physics \cite{7,8,9}. Moreover, BH thermodynamics is a branch of study that connects gravity and quantum mechanics and had been a major focus for researchers over the past thirty years, highlighting its importance \cite{10,11,12,13,14,15,16,17,18,19,20,21,22,23,24,25,25c,25a,25b}. Bekenstein and Hawking's 1960s and 1970s contributions laid the foundation for studies on BH as thermal systems, establishing laws (four), Hawking radiation, and BH entropy of BH thermodynamics \cite{26,27,28,29}. Recently, the thermodynamics of BH, had analyzed also through topological properties, had provided a deeper understanding of the intrinsic characteristics of space-time.

BHs are thermodynamic entities with a temperature equivalent to their gravity at the surface $\kappa$ and an entropy '$\mathcal{S}$' that equals around a quarter of the size $\mathcal{A}$ of the event horizon. Research on Kerr-AdS BH with changing cosmological constant $(\textsl{CC})$ shows $\mathbb{P}-\mathcal{V}$ criticality resembling traditional thermodynamic systems, with significant literature \cite{30,31} on various AdS BH. Sakharov \cite{26a} and Gliner \cite{26b} put forward avoiding singularities in the de-Sitter core matter, which Bardeen applied to propose the first solution for the static spherically symmetric regular BH. The literature \cite{26c,26d,26e} had examined the thermodynamics of the Schwarzschild de Sitter BH. This understanding necessitates the use of enthalpy to represent the mass of a BH, rather than internal energy, and interprets the negative $\textsl{CC}$ as an effective thermodynamic pressure, $\mathbb{P} = -\frac{\Lambda}{8 \pi}$ \cite{32,33,34,35,36,37,38}. Kastor \emph{et al}. \cite{38a} treated negative $\textsl{CC}$ as a thermodynamic pressure in the anti-de Sitter space-time. The thermodynamic volume $\mathcal{V}$, defined as $\mathcal{V}= (\frac{\partial~ H}{\partial \mathbb{P}})_{\mathcal{S}}$, is conjugate to $\mathbb{P}$. Notably, the temperature of a BH is influenced by the $\textsl{CC}$ and the BH radius, with the latter being intricately linked to $\mathcal{V}$. By reversing this relationship and applying it as an equation of state (EoS) in the BH framework, classical thermodynamic techniques can be employed to analyze the critical behavior of BH. Asymptotically, such BH tends towards a flat spatial configuration. Depending on the sign of $\Lambda$, Einstein's equations yield asymptotic solutions corresponding to either de-Sitter (for $\Lambda>0$) space or AdS space (for $\Lambda<0$). Felice and Tsujikawa \cite{38abc} explored several applications of a specific modified gravity theory, covering topics like cosmic inflation, dark energy, local gravity tests, cosmological perturbations, and solutions with spherical symmetry under both weak and strong gravitational fields.

A new method has been proposed by Glavan and Lin \cite{38b} to obtain a four-dimensional solution of the Einstein-Gauss-Bonnet gravity (EGBG) \cite{38c,38d}. It has been found that EGBG is a key component of Lovelock's theory, enhancing general relativity by incorporating higher derivative terms into the Einstein-Hilbert action \cite{38b}. Further, they found 4-dimensional nonzero limits of EGBG theory and its solution by rescaling the GBT by $\frac{1}{(D - 4)}$, where D is the space-time dimension. It has been observed that the Gauss-Bonnet term (GBT) emerges naturally as the first-order correction in the effective action of closed string theories, demonstrating a significant relationship \cite{39,40}. The groundbreaking research on the Gauss-Bonnet AdS BH solution has significantly advanced our understanding in this particular field, highlighting the importance of this theory and representing a significant breakthrough \cite{41,41a}. Studies on the thermodynamic aspects of GB BH in AdS backgrounds had been explored under different conditions, unveiling  a diverse selection of dynamic behaviors and unique characteristics. The analysis of the GBT primarily occurs in higher dimensions, revealing its impact on four-dimensional geometry. Derived from the low-energy aspect of heterotic string theory, the GBT enhances our comprehension of gravity on quantum scales \cite{42,43,44,45,46,47}. The inclusion of the GBT in this theory is particularly noteworthy for producing ghost-free models and yielding field equations that incorporate only the second-order derivatives of the metric, a property that distinguishes it from other modifications of general relativity \cite{48,49,50,51,52}. Nojiri et al. \cite{52abc} addresses the ghost issue, which is known to be plagued with ghost degrees of freedom in a particular class of GB gravity.

The GBT in BH thermodynamics may lead to observable signatures. It alters the relationship between entropy and area, impacting temperatures and heat capacities, and modifies Hawking radiation, potentially affecting evaporation rates and lifetimes. Additionally, the GBT changes the dimensions and shape of a BH shadow, influencing its appearance in observations. Ongoing and upcoming experiments can examine the GBT effects, with future X-ray and gamma-ray observatories studying quasinormal modes. Indirect investigations can focus on BH masses, spins, and accretion rates. Although the effects of the GBT may be modest and difficult to detect, continued research could provide valuable insights into modified gravity concepts.

In addition, it has been investigated about relativistic star structure, GB coupling $\alpha$, geodesic motion, shadow effects on BHs, and quasinormal modes in EGBG \cite{52a,52b}. They also had examined observational restrictions and BH quasinormal modes \cite{52c}. In Ref.~\cite{52d}, the study had looked at gravitational lensing by static and spherically symmetric BH in the strong deflection limit. The study in Ref.~\cite{52e} examined the gravitational instability of asymptotically flat, de-Sitter, and anti-de-Sitter BH in four-dimensional EGBG. Furthermore, the study explores the phase transition and thermodynamic behavior of BH, with additional research on four-dimensional EGBG in Refs.~\cite{52f,52g}. More studies on four-dimensional EGBG can be found in Refs.~\cite{52h,52i,52j,52k}.

The existing literature \cite{53,54} has delved into the intriguing properties of BH solutions in the context of EGBG. Specifically, the RN-AdS solutions, which had incorporated fictious fields within modified gravity theories \cite{54a,54b,54c,Sotiriou:2008rp,54d,54e,54f,54g,54h}, had offered a distinctive lens through which the interaction between gravity and exotic forms of matter can be understood. By analyzing these solutions, which result from the coupling of the electromagnetic field with a constant $\eta$ \cite{55}, we gained deeper insights into the characteristics of phantom energy, the fundamental nature of gravity, and their consequential roles in the thermodynamics of BH.
Furthermore, this study investigates the effects of thermal fluctuations within the framework of EGBG, employing an enhanced entropy formulation. Specifically, it examines key thermodynamic parameters such as Gibbs free energy (GibbsFE) and thermal stability, utilizing entropy modified by the GBT. Here, the given article examined the $\mathbb{P}-\mathcal{V}$ criticality of the RN-AdS BH within the EGBG.

In this work, we investigate the thermodynamic parameters, such as critical points and thermal stability, using GB-modified entropy. The horizon radius and the foundational principles of thermodynamics spur further interest in the thermal behavior of AdS BH. This article is organized as follows. The subsequent section delves into thermodynamic quantities and Gauss-Bonnet entropy as they pertain to the RN-AdS BH. Section \textbf{III} deals with the EoS and identifies critical points. Section \textbf{IV} is dedicated to the evaluation of thermodynamic stability. Finally, section \textbf{V} concludes with our main findings and remarks.

\section{Thermodynamic Quantities and Gauss-Bonnet Modified Entropy}
The EGBG represents a straightforward extension of Einstein's theory of gravity, incorporating higher-order curvature corrections, particularly relevant in five and six dimensions. Lovelock's theory of gravity \cite{57}, from which EGBG derives, is founded on three key elements: the (\textsl{CC}), the Einstein-Hilbert action, and the GBT \cite{58}.
The action for EGBG in D-dimensional space-time can be summed up as a result \cite{59}
\begin{eqnarray}\label{1}
\mathcal{I}= \frac{1}{16\pi}\int d^{D}x \sqrt{\textbf{g}}\left[\frac{1}{2 k^{D-2}}(\mathcal{R} -
2\Lambda + \alpha \mathcal{{\L}}_{GB}) + \mathcal{L}_\mathrm{matter}\right].
\end{eqnarray}
This equation uses $\mathcal{L}_\mathrm{matter}$ as the Lagrangian, $\textsl{k}^{D-2}$ as
the gravitational D-dimensional constant, and $\alpha$ as the GB coupled constant.
Furthermore, $\textbf{g}$ determines the metric tensor ($\textbf{g}_{\mu\nu})$. The Lagrangian is given by
\begin{eqnarray}\label{2}
\mathcal{{\L}}_{GB}=\mathcal{R}^{2}-4 \mathcal{R}^{\mu\nu}\mathcal{R}_{\mu\nu}+
\mathcal{R}^{\mu\nu\rho\sigma} \mathcal{R}_{\mu\nu\rho\sigma},
\end{eqnarray}
where $\mathcal{R}_{\mu\nu},~\mathcal{R}$, and $\mathcal{R}_{\mu\nu\rho\sigma}$ represent the Ricci tensor, scalar curvature, and Riemann curvature tensor, respectively, while $\Lambda$ denotes the $(\textsl{CC})$. The GBT emerges naturally from the lower-energy condition of the heterotic theory of strings \cite{60}. Juan \emph{et al.} \cite{61} introduced a novel type of regular BH solution for non-linear electrodynamics, incorporating a modified \textsl{CC}. The exploration of BH thermodynamics has been significantly advanced by the integration of quantum field theory in curved geometry, facilitating analogies between the event horizon of a BH and thermodynamic variables that involve the mass and temperature along entropy. This section aims to provide an overview of the thermodynamics associated with spherically symmetric AdS BH \cite{55}.
The line element of the metric is defined as \cite{55}
\begin{eqnarray}\label{3}
ds^{2} &=&\bold{\Delta( \textsl{r})} dt^{2} - [\bold{\Delta( \textsl{r})}]^{-1}  \textsl{dr}^{2} -  \textsl{r}^{2}(d \theta^{2} + \sin^ {2}\theta ~d \phi^{2})~,
\end{eqnarray}
where, the metric term $\bold{\Delta(\textsl{r})}$ is defined as follows
\begin{eqnarray}\label{4}
\bold{\Delta(\textsl{r}}) &=& 1- \frac{2 \mathcal{M}}{\bold{\textsl{r}}}-\frac{\Lambda ~ \bold{\textsl{r}}^2}{3} + \frac{\mathcal{Q}^2 ~ \eta}{\bold{\textsl{r}}^{2}}~,
\end{eqnarray}
here, the mass $\mathcal{M}$ is featured in the metric function $\bold{\Delta}$,
which characterizes the space-time geometry. The line element (4) has three horizons: $r_{+}$, $r_{-}$, and $r_{c}$. The event horizon is denoted by $r_{+}$, the Cauchy horizon by $r_{-}$, and the outer cosmological horizon by $r_{c}$. This metric function ensures the recovery of the Newtonian limit in the absence of the \textsl{CC} (i.e., $\Lambda = 0$), where the space-time becomes asymptotically Minkowskian. To examine the thermodynamic characteristics of this solution, it is necessary to define the mass $\mathcal{M}$ in terms of the charge $\mathcal{Q}$ and the radius of the event horizons $\textsl{r}_{+}$.
The space-time metric transitions to the well-known RN-AdS space-time metric, or anti-RN-AdS BH
space-time, when $\eta= 1$ and for case $\eta = -1$ with $\bold{\Lambda < 0}$.
The surface gravities at the horizons can be found by,
\begin{eqnarray}\label{5}
\kappa_{+} &=& \bigg|\frac{\mathcal{M}}{\textsl{r}_{+}^{2}} -  \frac{\Lambda ~ \textsl{r}^2_{+}}{3} + \frac{\mathcal{Q}^2 ~ \eta}{\textsl{r}^{2}_{+}}\bigg|~,
\end{eqnarray}
\begin{eqnarray}\label{6}
\kappa_{-} &=& -\bigg|\frac{\mathcal{M}}{\textsl{r}_{-}^{2}} -  \frac{\Lambda ~ \textsl{r}^2_{-}}{3} + \frac{\mathcal{Q}^2 ~ \eta}{\textsl{r}^{2}_{-}}\bigg|~,
\end{eqnarray}
\begin{eqnarray}\label{7}
\kappa_{c} &=& \bigg|\frac{\mathcal{M}}{\textsl{r}_{c}^{2}} -  \frac{\Lambda ~ \textsl{r}^2_{c}}{3} + \frac{\mathcal{Q}^2 ~ \eta}{\textsl{r}^{2}_{c}}\bigg|~,
\end{eqnarray}
\begin{center}
\begin{figure}[th!]
\epsfig{file=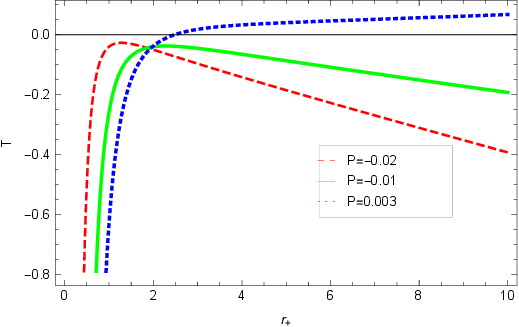, width=.65\linewidth, height=2.1in}
\caption{The behavior of $ \textsl{T}$ versus $\textsl{r}_{+}$ for a fixed value of $\eta = -1$, different values of $\mathcal{Q}=1, 2, 3$ and P. Each line in the plot corresponds to a specific value of the thermodynamic pressure P= -0.02 (pink dashed line), P= -0.01 (green solid line), and P= 0.003 (blue dotted line). }\label{f1}
\end{figure}
\end{center}
where the surface gravities of the event horizon, Cauchy horizon, and cosmological horizon are denoted by $\kappa_{+}$, $\kappa_{-}$, and $\kappa_{c}$, respectively.
Moreover, to find the mass $\mathcal{M}$ of the BH, we impose $\bold{\Delta(\textsl{r}}_{+})=0$ in Eq. (\ref{4}) to obtain
\begin{eqnarray}\label{8}
\mathcal{M} &=& \frac{ \textsl{r}_{+}}{2}\bigg(1-  \frac{\Lambda ~ \textsl{r}^2_{+}}{3} + \frac{\mathcal{Q}^2 ~ \eta}{\textsl{r}^{2}_{+}}\bigg)~.
\end{eqnarray}
Hawking, in a groundbreaking approach, developed the concept of BH temperature, now known as the Hawking temperature. He posited that BH emits thermal radiation, a phenomenon conceptualized as Hawking radiation. This insight was synergistic with Bekenstein's earlier proposal that BH should have an entropy. Integrating these ideas, Hawking proposed that BH indeed have a temperature, derived from their quantum emissions. The Hawking temperature for RN AdS BH can be calculated using the principle of surface gravity, as outlined below.
\begin{eqnarray}\label{9}
 \textsl{T} &=& \frac{\kappa}{2 ~\pi}.
\end{eqnarray}
The surface gravity ${\kappa}$ can be determined by applying the metric of the function $\bold{\Delta(\textsl{r}_{+})}$, that is presented by
\begin{eqnarray}\label{10}
\kappa = \frac{\bold{\Delta}~^{'}(\textsl{r}_{+})}{2}~.
\end{eqnarray}
Thus, $\frac{{\kappa}}{2 \pi}$ represents the exact temperature associated with a BH.
By using the formula $ \textsl{T} = \frac{\kappa}{2 ~\pi}$ and Eq. (\ref{4}).
One can also find the temperature of RN-AdS BH.
\begin{eqnarray}\label{11}
 \textsl{T} =\frac{\mathcal{M} }{2 ~\pi~ \textsl{r}^2_{+}} - \frac{ \mathcal{Q}^2 ~ \eta}{2 ~\textsl{r}^3_{+}~\pi} - \frac{\Lambda ~\textsl{r}_{+}}{6 ~ \pi}.
\end{eqnarray}
Inserting the value of $\mathcal{M}$ and using the relation $\mathbb{P} =\frac{-\Lambda}{8 \pi}$ in Eq. (\ref{11}), considered to be
\begin{eqnarray}\label{12}
 \textsl{T} =\frac{1}{4 ~ \pi~\textsl{r}_{+}}- \frac{ \mathcal{Q}^2 ~ \eta}{4 ~ \pi\textsl{r}^3_{+}} + 2~ \mathbb{P}~ \textsl{r}_{+}.
\end{eqnarray}
Alternatively, one can find the temperature as
\begin{eqnarray}\label{13}
 \textsl{T} = \bigg(\frac{\partial \mathcal{M}}{\partial \mathcal{S}}\bigg)_{\mathcal{Q}} =
\frac{\bigg(\frac{\partial \mathcal{M}}{\partial \textsl{r}_{+}}\bigg)_{\mathcal{Q}}}
{\bigg(\frac{\partial \mathcal{S}}{\partial \textsl{r}_{+}}\bigg)_{\mathcal{Q}}}.
\end{eqnarray}
\begin{center}
\begin{figure}[th!]
\epsfig{file=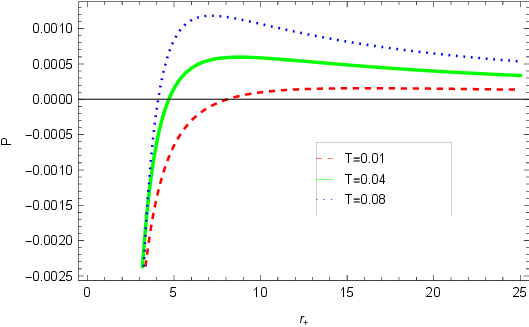, width=.65\linewidth, height=2.1in}
\caption{$\mathbb{P}$ versus $\textsl{r}_{+}$ for $\mathcal{Q}=1, 2, 3$ and $\eta = -1$ of RN-AdS BH.}\label{f2}
\end{figure}
\end{center}
Figure \textbf{\ref{f1}} displays three temperature curves as functions on the event horizon radius denoted by '$\textsl{r}_{+}$'. These curves illustrate the effect of varying the pressure values $P$ on the temperature behavior. As $\textsl{r}_{+}$ increases, the temperature generally decreases.
Notably, for $\textsl{r}_{+} > 2$, the temperature reaches a regime where $ \textsl{T} \leq 0$,for red and blue curves indicating non-physical or extremal conditions.
Figure \textbf{\ref{f2}} illustrates the relationship between pressure $\mathbb{P}$ and the event horizon $\textsl{r}_{+}$ for
different values of $ \textsl{T}$. As the horizon radius $\textsl{r}_{+}$ varies,
the pressure exhibits a transition from negative to positive values.
Specifically, at a temperature $ \textsl{T} = 0.08$, the pressure increases and attains positive values for larger values of $\textsl{r}_{+}$. When $ \textsl{T}=0.04$, the pressure increases and attains negative values
for smaller horizon radius values and positive for the $\textsl{r}_{+} \geq 3.5$.
For $ \textsl{T} =0.01$, the pressure increases for relatively small radius values at
$\textsl{r}_{+} \geq 3.5$. It is negative for $\textsl{r}_{+}\leq 9$ and positive for $\textsl{r}_{+} \geq 8.5$.
We note that pressure increases with larger horizon values $\textsl{r}_{+}$.
The volume of the BH leads
\begin{eqnarray}\label{14}
\mathcal{V}=\frac{4}{3}~\pi ~\textsl{r}_{+}^{3},\\\label{14}
\mathcal{V}= \bigg(\frac{\partial \mathcal{M}}{\partial \mathbb{P}}\bigg)_{\mathcal{S}, \mathcal{Q}}.
\end{eqnarray}
Moreover, the first thermodynamic law for RN-AdS BH is as follows
\begin{eqnarray}\label{15}
d \mathcal{M} =  \textsl{T} d\mathcal{S} - \eta \frac{\mathcal{Q}}{\textsl{r}_{+}} dq.
\end{eqnarray}
In the 1970s, theorists made the ground-breaking discovery that BH possess entropy \cite{62}, a revelation that prompted analogies between these space-time singularities and particle systems such as classical gases. It was Stephen Hawking's application of quantum mechanics that demonstrated BH emit radiation, akin to black bodies at a specific temperature \cite{63} a pivotal piece of evidence. This discovery resulted in the development and application of the laws (four) of thermodynamics to BH, ensuring their conceptual introduction into thermodynamic theory.
According to the Bekenstein-Hawking entropy, which was initially developed within the context of GR, a BH's entropy is proportional to the area of its event horizon ($\mathcal{S_{BH}} = \frac{\mathcal{A}}{4}$). However, when taking into account more comprehensive theories of gravity that incorporate higher order curvature factors, like the EGBT, this definition becomes inadequate because it is particular to GR. In addition, the Wald entropy is the proper and basic definition of BH entropy for these broader theories. The Noether charge connected to the gravitational action's diffeomorphism invariance yields the thermodynamic value known as the Wald entropy. For any stationary BH solution in any diffeomorphism invariant theory of gravity, it offers a strong, first principles definition.
The following is the general expression for Wald entropy:
\begin{eqnarray}\label{16}
\mathcal{S_{W}}= -2 \pi \int_{H} \bigg(\frac{\partial \mathcal{L}}{\partial \mathcal{R}_{\mu\nu\rho\sigma}}\bigg) \epsilon_{\mu\nu} \epsilon_{\rho\sigma}  d^{D-2} x,
\end{eqnarray}
In this case, $\epsilon_{\mu\nu}$ is the binormal to the horizon $H$, $\mathcal{R}_{\mu\nu\rho\sigma}$ is the Riemann tensor, and $\mathcal{L}$ is the theory's Lagrangian density. However, the Lagrangian contains a GBT that alters the connection between entropy and horizon area for the particular instance of the four dimensional EGB theory under consideration.
Thus, using the Hawking area law \cite{64},  the entropy $\mathcal{S}$ can be derived as $\mathcal{S} = \frac{\mathcal{A}}{4}$, where the area of the horizon $\mathcal{A}$ can be obtained by $4 \pi \textsl{r}^2_{+}$.
\begin{eqnarray}\label{16}
\mathcal{S}=\frac{\mathcal{A}}{4}+ 4\alpha \pi,
\end{eqnarray}
The generic statement reduces to the entropy formula utilized in this situation when the Wald entropy formalism is applied to the particular solution, while BH entropy is given by
\begin{eqnarray}\label{17}
\mathcal{S} =  \pi~\textsl{r}^{2}_{+} + 4 \alpha \pi.
\end{eqnarray}
This formula is a direct and justified result of the Wald entropy formalism rather than an arbitrary assumption. It makes it evident that the standard Bekenstein Hawking area law ($A=4 \pi~\textsl{r}^{2}_{+}$) has a correction term, $4 \pi~\alpha $. The EGB Lagrangian's higher order curvature terms are the immediate source of this extra term. In thermodynamics, entropy plays an important role between the macroscopic and microscopic worlds. In simple words, it measures a system's chaos, which BH possesses. There additionally exists a chemical potential, which is
\begin{eqnarray}\label{18}
\phi= \bigg(\frac{\partial \mathcal{M}}{\partial \mathcal{Q}}\bigg)_{\mathcal{S}}.
\end{eqnarray}
Which turns to
\begin{eqnarray}\label{19}
\phi &=&   \frac{\eta~ \mathcal{Q}}{\textsl{r}_{+}}.
\end{eqnarray}
The following relation can be used to compute the GibbsFE.
\begin{eqnarray}\label{20}
G= \mathcal{M}- \textsl{T}~\mathcal{S}.
\end{eqnarray}
The GibbsFE of a system at any given time is defined as the system's enthalpy minus the product of the system's temperature and entropy. GibbsFE serves as a measure of the thermodynamic potential when both pressure ($\mathbb{P}$) and temperature ($ \textsl{T}$) are held constant, effectively representing the work obtainable from a thermodynamic process.
\begin{eqnarray}\label{21}
G &=& \frac{\textsl{r}_{+}}{4} + \frac{\Lambda~ \textsl{r}^{3}_{+}}{6}+\frac{
3 ~ \mathcal{Q}^2 ~ \eta}{4 \textsl{r}_{+}} - \frac{4 \alpha}{\textsl{r}_{+}} + 4 ~\Lambda ~\alpha ~\textsl{r}_{+} + \frac{4 \alpha  \mathcal{Q}^2 ~ \eta}{\textsl{r}^{3}_{+}}.
\end{eqnarray}
\begin{center}
\begin{figure}[t]
\epsfig{file=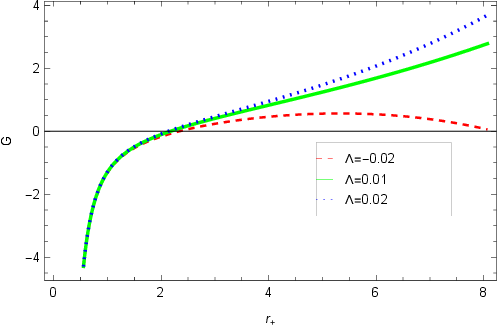, width=.65\linewidth, height=2.1in}
\caption{Function of $G$ versus $\textsl{r}_{+}$ for parameter values $\alpha=0.1$, $\mathcal{Q}=1$ and $\eta = -1$ of RN-AdS BH.}\label{f3}
\end{figure}
\end{center}
The GibbsFE ($G$) demonstrates a decrease as the horizon radius ($\textsl{r}_{+}$) increases, indicating potential global stability. This behavior is illustrated in figure \textbf{f3}, which plots $G$ against $\textsl{r}_{+}$ for various values of $\textsl{CC}$. Specifically, for $\Lambda = 0.01$ (solid curve) and $\Lambda = 0.02$ (dotted curve), $G$ shows an increasing trend for $\textsl{r}_{+}$ values greater than or equal to 0.5. For a negative $\textsl{CC}$ value ($\Lambda = -0.02$), the observed quantity increases over the interval $0.5 \leq \textsl{r}_{+} \leq 4$. Beyond this, when $\textsl{r}_{+} \geq 4$, it begins to show a decreasing trend. 

\section{Critical Points and EoS}

This section will examine the significance of the $\mathbb{P}-\nu$ relationship in the dynamics of the RN-AdS BH. The association of the $\textsl{CC}$ with thermodynamic pressure, represented by $\mathbb{P}=-\frac{\Lambda}{8 \pi}$, results in VdW-like behavior and first-order phase transitions. One of the most fascinating aspects of BH thermodynamics is the exploration of phase transitions and critical points. Researchers have developed new methodologies to identify these critical points and analyze phase transitions within the thermodynamic framework of BH. The VdW fluid model serves as the simplest, and most widely used model for a system of interacting particles undergoing a phase transitions. We now introduce the EoS, a thermodynamic equation that relates pressure ($\mathbb{P}$), temperature ($ \textsl{T}$), and volume ($\mathcal{V}$). In this context, `volume' pertains to a specific volume, diverging from the conventional usage of $\mathcal{V}$. Recent research suggests that the isotherms of Bardeen BH exhibit similarities to those of VdW fluids \cite{65}. By analyzing Eq. \ref{12}, we identify critical points using the EoS for $\mathbb{P}
= \mathbb{P}( \textsl{T}, \nu)$
which is in the form
\begin{eqnarray}\label{22}
\mathbb{P} = \frac{ \textsl{T}}{2~\textsl{r}_{+}}- \frac{1}{8 \pi~ \textsl{r}^{2}_{+}} + \frac{q^2 ~ \eta}{8 \pi ~\textsl{r}^{4}_{+}}.
\end{eqnarray}
This equation represents the thermodynamic EoS. Incorporating the Maxwell construction \cite{66} into the VdW theory describes phase coexistence along with ($\mathbb{P}$) versus volume ($\mathcal{V}$) isotherms, leading to a first-order transition. To link a specific volume $\nu$ to the BH horizon radius $\textsl{r}_{+}$, we use the VdW equation and an infinite series solution for the converse of the specific volume, i.e., $\nu = \frac{2}{(\textsl{r}_{+})^{-1}}$ \cite{67} \textendash \cite{69}. This introduces the concept of thermodynamic volume $\mathcal{V}$. By considering the EoS in terms of the specific volume $\nu$, we can explore
\begin{eqnarray}\label{23}
\nu = \frac{2}{(\textsl{r}_{+})^{-1}}.
\end{eqnarray}
To have a BH in the VdW, set $\mathbb{P} = \mathbb{P}_{\nu}$, then the EoS takes the form
\begin{eqnarray}\label{24}
\mathbb{P}_{\nu}=\frac{ \textsl{T}}{\nu} -  \frac{1}{2 \pi \nu^2} + \frac{2~ q^2 ~ \eta}{\pi \nu^{4}}.
\end{eqnarray}
\begin{center}
\begin{figure}[t]
\epsfig{file=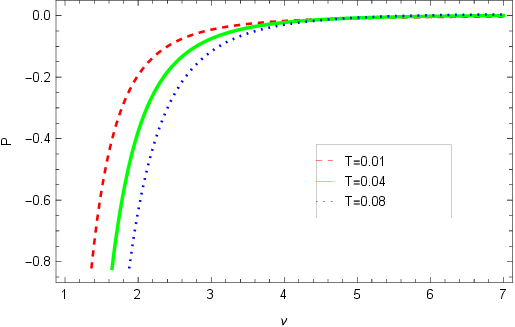, width=.65\linewidth, height=2.1in}
\caption{Function of $\textbf{P}$ versus $\nu$ for $\mathcal{Q}=1, 2, 3$ and $\eta = 0.1$ of RN-AdS BH.}\label{f5}
\end{figure}
\end{center}
Figure \textbf{\ref{f5}} illustrates the isotherms in a $\mathbb{P}-\nu$ plot, revealing temperature-dependent isothermal behavior. The isothermal curves demonstrate that above the critical temperature $ \textsl{T}_{c}$, the system behaves into an ``ideal gas state'', featuring a point of inflection. Below this critical temperature, the system transitions to a ``liquid-gas state''.
\begin{eqnarray}\label{25}
 \textsl{T}_{\nu}=\frac{1}{2 \pi \nu} +  \mathbb{P} \nu - \frac{2 \mathcal{Q}^2 ~ \eta}{\pi ~\nu^{3}}.
\end{eqnarray}
If there are any critical points, they are the inflection points on the isotherms, and therefore, they must satisfy the following conditions:
\begin{eqnarray}\label{26}
\frac{\partial \mathbb{P}}{\partial
\nu}|_{ \textsl{T}= \textsl{T}_{c},\nu =\nu_{c}}= 0,
\end{eqnarray}
\begin{eqnarray}\label{27}
\frac{\partial^{2}
\mathbb{P}}{\partial^{2} \nu}|_{ \textsl{T}= \textsl{T}_{c},\nu = \nu_{c}}=\frac{\partial}{\partial
\nu}\bigg(\frac{\partial \mathbb{P}}{\partial \nu}\bigg)=0.
\end{eqnarray}
Using Eqs. (\ref{24}) and (\ref{26}), we can find
\begin{eqnarray}\label{28}
\frac{\partial \mathbb{P}}{\partial \nu} &=& -\frac{ \textsl{T}_{c}}{\nu^2_{c}} + \frac{1}{\pi ~\nu^{3}_{c}}- \frac{8 ~ \eta~ \mathcal{Q}^2}{\pi ~\nu^{5}_{c} },
\end{eqnarray}
and using Eqs. (\ref{24}), (\ref{27}) and (\ref{28}), we have
\begin{eqnarray}\label{29}
\frac{\partial}{\partial \nu}\bigg(\frac{\partial \mathbb{P}}{\partial
\nu}\bigg)&=& - \frac{2  \textsl{T}_{c}}
{\nu^{3}_{c}} -
\frac{3}{\pi ~ \nu^{4}_{c}} +  \frac{40~\eta~ \mathcal{Q}^2}{\pi ~ \nu^{6}_{c}}.
\end{eqnarray}
We can find the critical temperature by comparing Eqs. (\ref{27}) and (\ref{29}).
\begin{eqnarray}\label{30}
 \textsl{T}_{c}=\frac{1}{\pi~ \nu_{c}} -  \frac{8~\eta~ \mathcal{Q}^2}{\pi~\nu^{3}_{c} }.
\end{eqnarray}
Putting in Eq. (\ref{24}), we get $\mathbb{P}_{c}$ as
\begin{eqnarray}\label{31}
\mathbb{P}_{c}&=&\frac{1}{2 \pi~\nu^2_{c}} -\frac{6 ~\eta~ \mathcal{Q}^2}{\pi~
\nu^{4}_{c}}.
\end{eqnarray}
Substituting Eq. (\ref{30}) in (\ref{29})
\begin{eqnarray}\label{32}
\frac{\partial^{2}\mathbb{P}}{\partial^{2}\nu}|_{ \textsl{T}= \textsl{T}_{c},\nu=\nu_{c}}&=&
\frac{-5}{\pi~\nu^{4}_{c}}-\frac{24 ~\eta~ \mathcal{Q}^2}{\pi~\nu^6_{c}}.
\end{eqnarray}
From Eqs. (\ref{30}), (\ref{31}) and (\ref{32}), we obtain critical points
$\nu_{c}$, $ \textsl{T}_{c}$ and $\mathbb{P}_{c}$,
which are given
\begin{eqnarray}\label{33}
\nu_{c}=\mathcal{Q} \sqrt{\frac{24 ~ \eta}{5}},~~ \bold{\textsl{T}_{c}}=\frac{2 \sqrt{5}}{3 ~\pi ~ \mathcal{Q} \sqrt{24~\eta~}},~~\bold{\mathbb{P}_{c}}= -\frac{5}{32 \pi~ \eta~ \mathcal{Q}^{2}}.
\end{eqnarray}
Critical exponents typically characterize behavior at critical points. The terms used are arranged as follows:
\begin{eqnarray}\label{34}
\tau= \frac{ \textsl{T}}{ \textsl{T}_{c}},  ~~p = \frac{\mathbb{P}}{\mathbb{P}_{c}},
~~\nu = \frac{\nu}{\nu_{c}}.
\end{eqnarray}
Critical factors are defined as follows:
\begin{eqnarray}\label{35}
 \hbar \propto \mid t {\mid}^{-\beta}, \textrm{C}_{\nu}\propto \mid t {\mid}^{{-\alpha}},
\end{eqnarray}
\begin{eqnarray}\label{36}
\kappa_{ \textsl{T}}\propto \mid t {\mid}^{-\gamma},~~~ \mathbb{P}-\mathbb{P}_{\nu}\propto \mid \nu_{3}-\nu_{c} {\mid }^{\delta}.
\end{eqnarray}
The study addresses heat behavior in a constant volume using $\textbf{a}$, $\beta$ for order parameter behavior, and $\gamma$ for isothermal compressibility behavior, $\kappa_{ \textsl{T}}$ and $\delta$
determines critical isotherm the behavior corresponding to
$ \textsl{T} =  \textsl{T}_{c}$.
All of such elements tend to be independent from one another \cite{69}.

\section{Thermal Stability Criteria}

Furthermore, our study aims to investigate thermal stability, a critical thermodynamic property that characterizes a system's response to small changes in thermodynamic parameters. In this context, specific heat emerges as a crucial quantity, offering insights into the system's thermal stability or instability. Specifically, a positive value of specific heat signals the stability of the system, while a negative value points to instability. To assess the thermal stability of the RN AdS BH, we traditionally rely on two metrics: isothermal compressibility and heat capacity. The formula for thermal compressibility is as follows:
\begin{eqnarray}\label{37}
\kappa_{ \textsl{T}}= -\frac{1}{\nu}~\frac{\partial
\nu}{\partial \mathbb{P}}\mid_{ \textsl{T}}.
\end{eqnarray}
The isothermal compressibility measures the variation of volume with pressure at constant temperature. For AdS BH, processes can be isochoric and isothermal when both the specific volume $\nu$ and entropy $\mathcal{S}$ remain constant, by the volume-entropy relation. Heat capacity can be viewed as:
\begin{eqnarray}\label{38}
\textit{C}=  \textsl{T} \bigg(\frac{\partial \mathcal{S}}{\partial
 \textsl{T}}\bigg)=  \textsl{T} \bigg(\frac{\partial \mathcal{S}}{\partial \textsl{r}_{+}}\bigg)\bigg(\frac{\partial  \textsl{T}}{\partial \textsl{r}_{+}}\bigg)^{-1}= 0,
\end{eqnarray}
from the above relation, we find
\begin{eqnarray}\label{39}
\textit{C}= \frac{2 \pi ~\textsl{r}^{2}_{+}\bigg( \Lambda ~\textsl{r}^{4}_{+} + \mathcal{Q}^2 ~
\eta - \textsl{r}^{2}_{+} \bigg)}{\bigg (\textsl{r}^{2}_{+} + \Lambda ~\textsl{r}^{4}_{+}
 - 3 \mathcal{Q}^2 ~ \eta \bigg)}.
\end{eqnarray}
This is the thermal stability of the RN-AdS BH. This understanding can be
combined with the constraint $\textit{C} \geq 0$ to study BH stability.
\begin{center}
\begin{figure}[t]
\epsfig{file=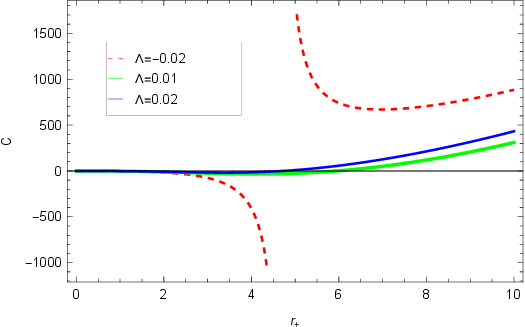, width=.65\linewidth, height=2.1in}
\caption{$\textit{C} as a function of $ versus $\textsl{r}_{+}$ for parameter values $\mathcal{Q}=1$ and $\eta = 0.1$ of RN-AdS BH.}\label{f6}
\end{figure}
\end{center}
Figure \textbf{\ref{f6}} displays the relationship between heat capacity and horizon radius. The metric's elements imply the stability criteria. Graphically representing the roots and divergences of heat capacity facilitates comprehension. The graphs of heat capacity exhibit positive and negative trends for various $\textsl{CC}$  values. A BH is locally unstable and stable for heat capacity $\textit{C} < 0$ and $\textit{C} > 0$, respectively. For $\Lambda=0.01$ and $\Lambda=0.02$, higher values of $\textsl{r}_{+}$ result in a positive heat capacity, indicating stability. Conversely, for $\Lambda = -0.02$, the heat capacity is negative, signaling instability, and its curve exhibits a discontinuity at $\textsl{r}_{+} = 5$. Beyond this point, the heat capacity becomes positive, indicating stability for $\textsl{r}_{+} \geq 5.5$, and reverts to negative, indicating instability, for $\textsl{r}_{+} \leq 5$. Utilizing the Hessian matrix, alongside Eqs. (\ref{4}) and (\ref{12}), and determining the matrix's determinant, it is found that the determinant diminishes due to the inherent properties of the matrix,
\begin{eqnarray}\label{40}
\bigg(\frac{d^{2} \Delta( \textsl{r}_{+})}{d \textsl{T}^{2}}\bigg)\bigg
(\frac{d^{2} \Delta( \textsl{r}_{+})}{d \eta^{2}}\bigg)=\bigg(\frac{d^{2}\Delta( \textsl{r}_{+})}
{d  \textsl{T} d \eta}
\bigg)\bigg(\frac{d^{2} \Delta( \textsl{r}_{+})}{d \eta d  \textsl{T}}\bigg).
\end{eqnarray}
\begin{eqnarray}\label{41}
\tau=\textsl {T} \tau(\mathcal{H}) = \tau_{1}+\tau_{2},
\end{eqnarray}
where
$\tau_{1}= \bigg(\frac{d^{2} \Delta( \textsl{r}_{+})}{d \textsl{T}^{2}}\bigg)$ and
$\tau_{2}= \bigg(\frac{d^{2} \Delta( \textsl{r}_{+})}{d  {\eta}^{2}}\bigg)$.
Now, the necessary condition for stability is $\tau \geq 0$.
\begin{eqnarray}\label{42}
\tau = \frac{\eta}{\textsl{r}_{+}} - 2 \pi~ \Lambda~ \textsl{r}_{+} - 2 \pi~ \frac{\eta~ \mathcal{Q}}{\textsl{r}^{3}_{+}}.
\end{eqnarray}
\begin{center}
\begin{figure}[t]
\epsfig{file=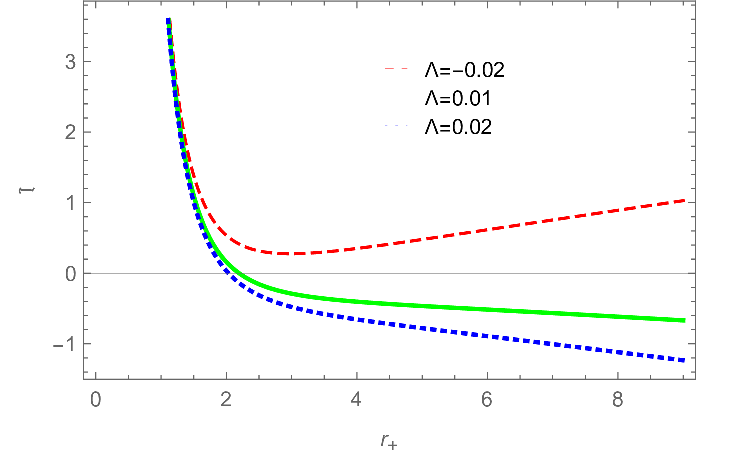, width=.65\linewidth, height=2.1in}
\caption{Function of $\tau$ against $\textsl{r}_{+}$ with parameter values $\eta = -1$ and $\mathcal{Q}=1$.}\label{f7}
\end{figure}
\end{center}
Figure \textbf{\ref{f7}} shows how $\tau$ varies with respect to $\textsl{r}_{+}$ of various $\Lambda$ values. The stability holds
for $\tau=0$, as $\tau \geq 0$ is required for the Hessian matrix's trace to be stable. For positive values of $\Lambda = 0.01$ and $\Lambda = 0.02$ the curve behavior shows $\tau \leq 0$, however for the negative value $\Lambda = -0.02$, the curve behavior
indicates $\tau \geq 0$ with the dashed curve demonstrating the conditions for stability according to the trace of the Hessian matrix.

\section{Conclusions}

This article introduces the metric known as the RN-AdS BH and provides a theoretical overview of Extended Gravity and its terms. Notably, the $\textsl{CC}$ is treated as a fluctuating quantity, manifesting as thermodynamic pressure where $\mathbb{P} =\frac{-\Lambda}{8 \pi}$, thereby linking $\Lambda$ with the proportionality of thermodynamic pressure. This represents a shift from treating the $\textsl{CC}$  as a mere constant to viewing it as a variable parameter within the system. Furthermore, we have employed a modified entropy formula, $\mathcal{S}=\pi~ (\textsl{r}_{+}^{2} +
4 ~\alpha)$, where $\alpha$ is referred to as a constant known as the Gauss-Bonnet constant, introducing new entropy considerations within this framework. Additionally, the GibbsFE pertinent to the GBT has been derived. In exploring the thermodynamics of this charged metric, an EoS $\mathbb{P} = (\mathbb{P},  \textsl{T})$ was utilized to analyze phase transitions within the BH context.

Applying the VdW equations $\frac{\partial}{\partial \nu} (\frac{\partial \mathbb{P}}{\partial \nu})$ and
$(\frac{\partial \mathbb{P}} {\partial \nu})$, we determined the important critical points $ \textsl{T}_{c}$, $\mathbb{P}_{c}$,
and $\nu_{c}$, as well as the critical exponents. We used modified corrected entropy $\mathcal{S}= \pi(\textsl{r}_{+}^2 + 4\alpha)$
to implement modified GibbsFE under GBT. We displayed the thermodynamic quantities pressure, temperature, GibbsFE, and heat capacity versus horizon radius. The graphs of $\mathbb{P}-\textsl{r}_{+}$, $ \textsl{T}-\textsl{r}_{+}$, $G-\textsl{r}_{+}$ and $\textit{C}$ show that the GibbsFE and $\textsl{T}$ increase with horizon radius. The figures depict three pressure curves through the horizon radius, representing isotherms that vary with temperature. In this present work, we have concluded some results of the thermal properties of the given metric which is RN-AdS BH. The temperature against the horizon radius is decreasing with larger values of the horizon radius. We can see that as going far away from the horizon of the RN-AdS BH, the temperature is going decreasing. Pressure is higher for $3\leq \textsl{r}_{+}\leq 5$ and after that it is decreased for greater values of $\textsl{r}_{+}$.

The local and global stability of RN-AdS BH checked through GibbsFE and heat capacity $\textit{C}$~ by choosing parameter values such that $\alpha = 0.1$, $\eta = -1$ and different values of $\mathcal{Q} = 1, 2, 3$ indicates that the GibbsFE is increasing for $\textsl{r}_{+}\geq 2$.
The equation of state in the $\mathbb{P}-\nu$ plot identifies critical points, indicating the system is in an `ideal gas state' when the temperature is above the critical temperature while below the critical temperature such that $\textsl{T} \leq \textsl{T}_{c}$, system behaved as ``liquid gas state". We also tested the thermal stability of RN-AdS BH under two conditions: isothermal compressibility $\kappa_{\textsl{T}}$ and heat capacity $\textit{C}$ and analyzed the behavior of the thermal stability. The heat capacity varies between negative and positive values depending on $\textsl{r}_{+}$. The RN-AdS BH is stable when $\textit{C}\geq 0$. Lastly, we employed the Hessian matrix to evaluate stability, adhering to the condition $\tau \geq 0$, and subsequently determined the trace of the Hessian matrix.

This work offers a fresh viewpoint beyond conventional treatments in General Relativity by using a modified entropy formalism originating from the GBT, so building upon previous research in BH thermodynamics. The derivation of the GibbsFE under this modified entropy and the subsequent analysis of critical points and thermal stability provide new insights into the impact of higher-curvature corrections on the phase structure of RN-AdS BHs. The GBT is a particular combination of curvature terms that appear in the action of some gravitational theories. These terms are quadratic in the Riemann tensor. Although this term can contribute dynamically in higher dimensions, it is topological in four dimensions. The following physical interpretations relate to the GB coupling constant $(\alpha)$ in the GB theory. The value of curvature correction can be controlled by the symbol $\alpha$. A value greater than zero denotes deviation from the standard Einstein gravity. Higher-order curvature terms become significant at high energies or in locations with substantial curvature, and their effects are controlled by the GB coupling constant. BH entropy, which is correlated with surface area, is altered by $\alpha$. Curvature singularities are less severe when $\alpha$ can control them. The amount that ``higher-order curvature" terms influence space-time geometry can be visualized as being represented by $\alpha$. Our findings demonstrate that the GB coupling leads to a richer phase diagram and modifies the conditions for local and global stability, as evidenced by the behavior of the GibbsFE and heat capacity. To sum up, the strength of curvature corrections is represented by the GB coupling constant $\alpha$, which has an impact on cosmology, gravitational waves, curvature effects, and BH features. Its value influences the stability and thermodynamic properties of the RN-AdS BH. For this article, the $\alpha$ term is particularly used in the modified corrected entropy. The thermodynamical properties like GibbsFE have the GB coupling constant $\alpha$ term that indicates the stability of the BH.~ For the given values, the GibbsFE indicates the global stability of the BH and $\alpha$ value does not affect the stability (it remains stable).\\

This work opens up a wide range of potential future research avenues that could lead to a greater comprehension of gravitational physics. Using the GBT framework to expand this study to higher-dimensional RN-AdS BHs is one convincing approach. Examining the effects of spacetime dimensionality on thermodynamic stability and critical phenomena, especially with the changed entropy, may provide new information. Furthermore, careful consideration should be given to examining the consequences of this altered entropy for the information paradox and the remnants of BHs. Examining the holographic duals of these BH solutions is another important area. The microscopic degrees of freedom of gravity and the nature of quantum gravity may be clarified by comprehending the boundary field theory corresponding to the core RN-AdS BH with modified entropy and GBT corrections. Investigating how these changes affect gravitational lensing and BH shadows may also be beneficial, as they might provide observable signs to constrain the parameter $\alpha$. This research has the potential to increase our understanding of gravity in extreme regimes and its relationship to quantum field theory, as well as to correlate theoretical studies of BH thermodynamics with observed occurrences. In the future study, we will investigate the thermodynamics of some other well-known BH with the modified entropy emphasizing the phase transitions under GBT.\\

\noindent  {\bf Acknowledgement:} The work of KB was supported by the JSPS KAKENHI Grant Numbers 21K03547, 23KF0008, 24KF0100 and Competitive Research Funds for Fukushima University Faculty (25RK011). The work by BA was supported by Researchers Supporting Project number: ORF2025R650, King Saud University,
Riyadh, Saudi Arabia.\\

\noindent {\bf Conflict of Interest:} The authors declare no conflict of interest.\\

\noindent  {\bf Data Availability Statement:}
This manuscript has no associated data or the data will not be deposited. [Authors comment: This manuscript
contains no associated data.]


\begin{thebibliography}{}

\bibitem{padmanabhan2007}
T.~Padmanabhan,
Gen. Rel. Grav. \textbf{40} (2008), 529-564.

\bibitem{1}  A.~Einstein, Ann. Phys. \textbf{14} (2005), 517.

\bibitem{2}  K. Schwarzschild, Math. Phys. \textbf{189} (1916), 9905030.

\bibitem{3}  H. Reissner, Ann. Phys. \textbf{59} (1916), 106.

\bibitem{4}  G. Nordstr\"{o}m, Ann. Phys. \textbf{20} (1918), 1238.

\bibitem{5}  R. P. Kerr, Phys. Rev. Lett. \textbf{11} (1963), 237.

\bibitem{6}  E. T. Newman, R. Couch, K. Chinnapared et al, J. Math. Phys. \textbf{06} (1965), 918.

\bibitem{7}  S. Gunasekaran, R. B. Mann and D. Kubiznak, J. High Energy Phys. \textbf{11} (2012), 110.

\bibitem{8}  S. W. Wei, P. Cheng and Y. X. Liu, Phys. Rev. D \textbf{93} (2016), 084015.

\bibitem{9}  N. Altamirano, D. Kubiznk, R. B. Mann and Z. Sherkatghanad, Class. Quantum Grav. \textbf{31} (2014), 042001.

\bibitem{10}  R. V. Pound, G. A. Rebka, Phys. Rev. Lett. \textbf{04} (1960), 337.

\bibitem{11}  A. Eckart, R. Genzel, Nature \textbf{383} (1996), 415.

\bibitem{12}  B. P. Abbott et. al, Phys. Rev. Lett. \textbf{116} (2016), 061102.
\bibitem{13}  M. M. Caldarelli, G. Cognola, and D. Klemm, Class. Quantum Grav. \textbf{17} (2000), 399.

\bibitem{14}  D. N. Page, New. J. Phys. \textbf{07} (2005), 203.

\bibitem{15}  T. Jacobson and A. C. Wall, Found. Phys. \textbf{40} (2010), 1076.

\bibitem{16}  B. C. Zhang, Phys. Rev. D \textbf{92} (2015), 081501.

\bibitem{17}  M. Appels, R. Gregory, and D. Kubizbnak, Phys. Rev. Rett. \textbf{117} (2016), 131303.

\bibitem{18}  M. Astorino, Phys. Rev. D \textbf{95} (2017), 064007.

\bibitem{19}  M. Dehghani, Phys. Lett. B \textbf{785} (2018), 274.

\bibitem{20}  C. H. Bayraktar, Eur. Phys. J. Plus. \textbf{133} (2018), 377.

\bibitem{21}  S. Q. Hu, Y. C. Ong, and D. N. Page, Phys. Rev. D \textbf{100} (2019), 104022.

\bibitem{22}  Y. Yao, M. S. Hou, and Y. C. Ong, Eur. Phys. J. C \textbf{79} (2019), 513.

\bibitem{23}  P. Krtous, and A. Zelnikov, J. High. Energy Phys. \textbf{2020} (2020), 164.

\bibitem{24}  M. Sharif, and H. S. Nawaz, Chin. Phys. C \textbf{67} (2020), 193.

\bibitem{25}  R. Gregory, Z. L. Lim, and A. Scoins, Front. Phys. \textbf{09} (2021), 666041.

\bibitem{25c} C. Singha, Gen. Relativ. Gravit. \textbf{54} (2022), 38.

\bibitem{25a}  S. Khan, A. Adeel, and Z. Yousaf, Eur. Phys. J. C \textbf{84} (2024), 572.

\bibitem{25b}  Z.Yousaf, K. Bamba, B. Almutairi, S. Khan, and M. Z. Bhatti, Class. Quantum Grav. \textbf{41} (2024), 175001.

\bibitem{26}  S. W. Hawking, Nature \textbf{248} (1974), 30-31.

\bibitem{26a}  A. D. Sakharov, Teor. Fiz. \textbf{49} (1966), 345.

\bibitem{26b}  E. B. Gliner, Sov. Phys. JETP \textbf{22} (1966), 241.

\bibitem{26c}  Y. Sekiwa, Phys. Rev. D \textbf{73} (2006), 084009.

\bibitem{26d}  E. B. Gliner, Sov. Phys. Phys. Rev. D \textbf{67} (2003), 104024.

\bibitem{26e}  M. Urano, A. Tomimatsu and H. Saida,  Class. Quantum Grav \textbf{26} (2009), 105010.

\bibitem{26f}  H. H. Zhao, L. C. Zhang, M. S. Ma and R. Zhao, Phys. Rev. D \textbf{90} (2014), 064018.

\bibitem{27}  S. W. Hawking, Commun. Math. Phys. \textbf{43} (1975), 199.

\bibitem{28}  R. G. Cai, Phys. Rev. D \textbf{65} (2002), 084014.

\bibitem{29}  D. J. Gross, E. Witten, Nucl. Phys. B \textbf{1} (1986), 277.

\bibitem{30}  N. Altamirano, D. Kubiznak and R. B. Mann, Phys. Rev. D \textbf{88} (2013), 101502.

\bibitem{31}  S. J. Yang, R. Zhou, S. W. Wei and Y. X. Liu, Phys. Rev. D  \textbf{105} (2022), 084030.

\bibitem{32}  C. Niu, Y. Tian and X. N. Wu, Phys. Rev. D \textbf{85} (2012), 024017.

\bibitem{33} S. H. Hendi, S. Panahiyan, and B. E Panah, Int. J. Mod. Phys. D \textbf{25} (2016), 1650010.

\bibitem{34} S. H. Hendi, S. Panahiyan, and B. E Panah, J. High Energy Phys. \textbf{1601} (2016), 129.

\bibitem{35} S. H. Hendi, S. Panahiyan, and B. E. Panah, Prog. Theor. Exp. Phys. \textbf{2015} (2015), 103.

\bibitem{36} S. H. Hendi, S. Panahiyan, B. E. Panah, and M. Momennia,  Ann. Phys. \textbf{528} (2016), 819.

\bibitem{37} S. D. P. Majumdar, and R. K. Bhaduri, Class. Quantum Grav. \textbf{19} (2002), 2355.

\bibitem{38} D. G. Boulware and S. Deser,  Phys. Rev. Lett. \textbf{55} (1985), 2656.

\bibitem{38a} D. Kastor, S. Ray and J. Traschen, Class. Quantm Grav. \textbf{26} (2009), 195011.

\bibitem{38abc} A. De Felice and S. Tsujikawa, Living Rev. Rel. \textbf{13} (2010) 3.

\bibitem{38b} D. Glavan and C. Lin,  Phys. Rev. Lett. \textbf{124} (2020), 081301.

\bibitem{38c}  H. L\"{u} and Y. Pang,  Phys. Lett. B \textbf{809} (2020), 135717.

\bibitem{38d}   T. Kobayashi,  J. Cosmol. Astropart. Phys.  \textbf{07} (2020), 013.

\bibitem{39}  R. R. Metsaev, A. A. Tseytlin, Phys. Lett. B \textbf{385} (1987), 293.

\bibitem{40}  M. C. Bento, O. Bertolami, Phys. Lett. B \textbf{198} (1996), 368.

\bibitem{41}  D. J. Gross, J. H. Sloan, Nucl. Phys. B \textbf{41} (1987), 291.

\bibitem{41a} S. Capozziello and G. G. L. Nashed, Class. Quantum Grav. \textbf{40} (2023) 205023.

\bibitem{42}  B. Zumino, Phys. Rep. \textbf{109} (1986), 137.

\bibitem{43}  R. C. Myers, Nucl. Phys. B \textbf{701} (1987), 289.

\bibitem{44}  C. G. Callan, R. C. Myers and M. J. Perry, Nucl. Phys. B \textbf{388} (2002), 621.

\bibitem{45}  Y. M. Cho, I. P. Neupane and P. S. Wesson, Adv. Theor. Math. Phys. \textbf{02} (1998), 231.

\bibitem{46}  D. G. Boulware, S. Deser: Phys. Rev. Lett. \textbf{55} (1985), 2656.

\bibitem{47}  D. Wiltshire: Phys. Rev. D \textbf{38} (1988), 2445.

\bibitem{48}  R. G. Cai, Q. Guo: Phys. Rev. D \textbf{69} (2004), 104025.

\bibitem{49}  A. Barrau, J. Grain, S. O. Alexeyev: Phys. Lett. B \textbf{584} (2004), 114.

\bibitem{50}  G. Dotti, J. Oliva, R. Troncoso: Phys. Rev. D \textbf{76} (2007), 064038.

\bibitem{51}  C. Charmousis, Lect. Notes Phys. \textbf{299} (2009), 769.

\bibitem{52}  S. H. Hendi, B. E. Panah, Phys. Lett. B \textbf{77} (2010), 684.

\bibitem{52abc} S. Nojiri, S.D. Odintsov, V.K. Oikonomou, Arkady A. Popov, Nuc. Phys. B \textbf{973} (2021), 115617.


\bibitem{52a}  M. Guo and P. C. Li, Eur. Phys. J. C  \textbf{80} (2020), 588.

\bibitem{52b}  R. Roy and S. Chakrabarti, Phys. Rev. D \textbf{102} (2020), 024059.

\bibitem{52c}  M. S. Churilova, Phys. Dark. Univ  \textbf{31} (2021), 100748.

\bibitem{52d}  R. Kumar, S. U. Islam and S. G. Ghosh,  Eur. Phys. J. C \textbf{80} (2020), 1128.

\bibitem{52e}  R. A. Konoplya and A. Zhidenko,  Phys. Dark. Universe  \textbf{30} (2020), 100697.

\bibitem{52f} S. A. Hosseini, Phys. Dark. Universe \textbf{31} (2021), 100776.

\bibitem{52g} D. V. Singh and S. Siwach, Phys. Lett. B \textbf{808} (2020), 135658.

\bibitem{52h} C. Y. Zhang, P. C. Li and M. Guo, Eur. Phys. J. C  \textbf{80} (2020), 874.

\bibitem{52i} R. A. Konoplya and A. Zhidenko, Phys. Rev. D \textbf{102} (2020), 064004.

\bibitem{52j} B. Eslam Panah, K. Jafarzade and S. H. Hendi, Nucl. Phys. B \textbf{961} (2020), 115269.

\bibitem{52k} X. H. Ge and S. J. Sin, Eur. Phys. J. C \textbf{80} (2020), 695.

\bibitem{53}  R. G. Cai, L. M. Cao, L. Li and R. Q. Yang, J. High. Energy Phys. \textbf{09} (2013), 005.

\bibitem{54}  S. H. Hendi, S. Panahiyan and E. Mahmoudi, Phys. Rev. D \textbf{74} (2014), 3079.

\bibitem{54a} T.~Padmanabhan, Gen. Rel. Grav. \textbf{40} (2008), 529-564.

\bibitem{54b} R.~Durrer and R.~Maartens, Gen. Rel. Grav. \textbf{40} (2008), 301-328.

\bibitem{54c} S.~Nojiri and S.~D.~Odintsov, Int. J. Geom. Meth. Mod. Phys. \textbf{4} (2007), 115-146.

\bibitem{Sotiriou:2008rp}
T.~P.~Sotiriou and V.~Faraoni,
Rev. Mod. Phys. \textbf{82}, 451-497 (2010).

\bibitem{54d} S.~Nojiri and S.~D.~Odintsov, Phys. Rep. \textbf{505} (2011), 59-144.

\bibitem{54e} S.~Nojiri, S.~D.~Odintsov and V.~K.~Oikonomou, Phys. Rep.\textbf{692} (2017), 1-104.

\bibitem{54f} S.~Capozziello and M.~De Laurentis, Phys. Rep.\textbf{509} (2011), 167-321.

\bibitem{54g} T.~Clifton, P.~G.~Ferreira, A.~Padilla and C.~Skordis, Phys. Rep.\textbf{513} (2012), 1-189.

\bibitem{54h} J.~de Haro, S.~Nojiri, S.~D.~Odintsov, V.~K.~Oikonomou and S.~Pan, Phys. Rep.\textbf{1034} (2023), 1-114.

\bibitem{55} D. F. Jardim, M. E. Rodrigues and M. J. S. Houndjo, Eur. Phys. J. Plus \textbf{127} (2012), 123.

\bibitem{56} A. Jawad, G. Abbas, I. Siddique and G. Mustafa, Eur. Phys. J. Plus \textbf{137} (2022), 284.

\bibitem{57} D. Lovelock, J. Math. Phys. \textbf{12} (1971), 498.

\bibitem{58} K. Izumi, Phys. Rev. D \textbf{90} (2014), 044037.

\bibitem{59} B. Zwiebach, Phys. Lett. B \textbf{156} (1985), 315.

\bibitem{60} A. M. Ghezelbash, J. High Energy Phys. \textbf{08} (2009), 045.

\bibitem{61} M. W. Juan, R. G. Cai and S. Keng, Commun. Theor. Phys. \textbf{46} (2006), 453.

\bibitem{62} B. P. Abbott et. al, Phys. Rev. Lett. \textbf{116} (2016), 061102.

\bibitem{63} J. D. Bekenstein, Phys. Rev. D \textbf{7} (1973), 2333-2346.

\bibitem{64} A. G. Tzikas, Phys. Lett. B \textbf{788} (2019), 219.

\bibitem{65} J. C. Maxwell, Nature \textbf{11} (1875), 357-359.

\bibitem{66} G. Q. Li, Phys. Lett. B \textbf{735} (2014), 256.

\bibitem{67} J. X. Mo, W. B. Liu, Phys. Lett. B \textbf{727} (2013).

\bibitem{68} R. G. Cai, L. Li, L. F. Li and R. K. Su, J. High Energy Phys. \textbf{06} (2013) 063.

\bibitem{69} S. W. Wei, Y. X. Liu, Phys. Lett. B \textbf{101}(2020), 104018.




%
%
%


%
%
%
%
%
%
%
%
%

%
%
%
%

%
%
%



%
%



\end{thebibliography}
\end{document}